\documentstyle[eqsecnum,preprint,aps]{revtex}
\begin{document}
\draft
\title{\bf Scaling Behavior
and Universality near the Quantum Hall Transition}
\author{K. Ziegler}
\address{Max-Planck-Institut f\"ur Physik Komplexer Systeme,
Au\ss enstelle Stuttgart, Postfach 800665, D-70506 Stuttgart,
Germany}
\date{\today}
\maketitle 
\begin{abstract}
A two-dimensional lattice system of non-interacting electrons in a 
homogeneous magnetic field with half a flux quantum per plaquette and a random
potential is considered. For the large scale behavior a supersymmetric theory
with collective fields is constructed and studied within saddle point
approximation and fluctuations. The model is characterized by a broken
supersymmetry indicating that only the fermion collective field becomes
delocalized whereas the boson field is exponentially localized. Power counting
for the fluctuation terms suggests that the interactions between delocalized
fluctuations are irrelevant. Several quasi--scaling regimes, separated by
large cross--over lengths, are found with effective exponents $\nu$ for the
localization length $\xi_l$. In the asymptotic regime there is $\nu=1/2$ in
agreement with an earlier calculation of Affleck and one by Ludwig et al. for
finite density of states. The effective exponent, relevant for physical
system, is $\nu=1$ where the coefficient of $\xi_l$ is growing with randomness.
This is in agreement with recent high precision measurements on Si MOSFET and
AlGaAs/GaAs samples.
\end{abstract}

\section{Introduction}

The transition between quantum Hall plateaux in a two-dimensional electron
gas is characterized by a divergent localization length $\xi_l$
with a critical exponent $\nu$ and a non-zero longitudinal conductivity
$\sigma_{xx}$. $\xi_l$ is finite and $\sigma_{xx}$ is zero inside the Hall
plateaux whereas $\sigma_{xy}$ is a constant.

A direct measurement of the localization length exponent in a AlGaAs/GaAs
sample by Koch et al. \cite{koch} gave a value for $\nu$ very close to $7/3$.
Recent high precision measurements on Si MOSFET \cite{dol1,dol2} and
AlGaAs/GaAs samples \cite{sha}, however, indicate that $\xi_l$ diverges with
the electron density $n$ like $\approx b_n(n_c-n)^{-1}$ or with the magnetic
field $H$ like $\approx b_H(H_c-H)^{-1}$, where the quantum Hall transition
(QHT) is at $n=n_c$ or $H=H_c$, respectively. The exponent $\nu\approx1$
appears to
be almost independent of the material or the Hall plateaux. On the other hand,
the coefficients $b_n$, $b_H$ are sensitive to disorder: they {\it increase}
with increasing disorder \cite{dol2,sha}. This is a remarkable observation
because in the scaling theory of Anderson localization \cite{LR} the
coefficient is related to the mean free path. That means it would 
{\it decrease}
with increasing disorder. The observation of $\nu\approx1$ is in sharp contrast
to the experiment by Koch et al. The disagreement was explained in \cite{dol1}
by insufficient sample size in the earlier experiment.

The localization length scale near the QHT was also
studied intensively in a number of numerical simulations using the network
model \cite{chalk,lee} and the lowest Landau level approximation
\cite{huck,huo}. These calculations agree on the result that the critical
exponent is $\nu\approx7/3$.

Concerning the exponent $\nu$ there is a calculation by Affleck \cite{affleck}
based on the $U(2n)/U(n)\times U(n)$
nonlinear sigma model with topological term in the replica limit
$n\to0$ \cite{pruisken}. He obtains $\nu=1/2$.
The same value was found for Dirac fermions with random vector potential if
the average density is finite \cite{ludwigetal}.
 
A reason for the impressive agreement of the numerical calculations, on the
one hand, and the disagreement between experiments, numerical and analytic
calculations on the other hand could be the sensitivity of the QHT
to the size of the system and to the type of disorder. In particular, it
may be related to the existence of a large characteristic length scale
depending on disorder. The existence of such a typical scale is also
indicated by the numerical results due to the fact that there is a cross-over
from the pure network model ($\nu=1$) to the random network model
($\nu\approx 7/3$).

The purpose of this article is to investigate the role of disorder induced
length scales in a tight--binding model with strong magnetic field. The
work is based on an effective supersymmetric field theory for Dirac fermions
with a random mass which enables us to study large scale properties.  
The main results are:\\
1. Spontaneous breaking of the supersymmetry. I.e., the effective field theory
for the QHT is {\it not} a nonlinear sigma model.\\
2. The scaling behavior of the localization length depends on the 
characteristic scale $\exp(\pi/g)$, where $g$ is the strength of disorder:
If $\xi_l\ll \exp(\pi/g)$ the effective exponent is $\nu=1$ whereas for
$\xi_l\gg \exp(\pi/g)$ the exponent is $\nu=1/2$.\\
3. There is a universal value for the conductivity $\sigma_{xx}=e^2/h\pi$.\\

The article is organized as follows: After the definition of the model in
Sect. II an effective field theory is constructed for the averaged Green's
functions (Sect. III).
This field theory includes the description of
the conductivity according to Kubo's formula.
Then a collective field representation is introduced in order to cover
symmetry breaking effects (Sect. III.A). The latter are discussed using a
saddle point approximation for the collective field (Sect. III.B). Gaussian
fluctuations around the saddle points and corrections to Gaussian fluctuations
are studied in Sects. III.C and III.D, respectively. Finally, the localization
length (Sect. IV) and the conductivity (Sect. V) are evaluated.

\section{The Model}

A lattice model is considered in this article, stressing the
universality in terms of the electron density and the magnetic field
observed in the experiment, to study the asymptotic behavior of the
localization length near the QHT.
Starting point is a microscopic model for non-interacting electrons on a
regular lattice in a homogeneous magnetic field. Disorder enters only through
a random potential on the lattice. This choice guarantees that the disorder
does not affect the homogeneity of the magnetic field.
The model is defined by the tight-binding Hamiltonian on a square lattice
with magnetic flux $\phi=B a^2$, where $a$ is the lattice constant and $B$
the homogeneous external magnetic field. There is nearest neighbor hopping
with rate $t$ and next nearest neighbor hopping with rate $t'$. 
The Hamiltonian reads in Landau gauge
\begin{eqnarray}
H & = & -\sum_r[te^{2i\pi Bay/\phi_0}c(r)c^\dagger(r+e_x)+tc(r)c^\dagger(r+e_y)
+t'e^{2i\pi Ba(y\pm{1\over2})/\phi_0}c(r)c^\dagger(r+e_x\pm e_y)
\nonumber\\
& & +h.c.]+\sum_rV(r)c(r) c^\dagger(r).
\label{hamiltonian}
\end{eqnarray}
$e_{x,y}$ are lattice unit vectors, and $c^\dagger$ and $c$ are fermion
creation and annihilation operators, respectively.
$V(r)$ is a random potential representing disorder on the lattice.
Without disorder, i.e., for $V(r)=0$, this model was discussed extensively in
the literature \cite{thou,hats,osh}. A central result is the occurence of
electron bands with a quantized Hall conductivity in each band.
Gaps can be created in the model, for instance, by choosing a staggered
potential $V(r)=(-1)^{r_1+r_2}\mu$ in the Hamiltonian (\ref{hamiltonian})
\cite{ludwigetal}.
By varying the staggered chemical potential one varies the concentration of
electrons in the system.
There are other methods to create gaps in a tigh-binding model. E.g.,
one could vary the magnetic field. This, however, would lead to a more
complicated situation because the corresponding vector potential depends
on space. In general, the relevant parameter for the quantum Hall transition
is the filling factor $n\Phi_0/B$. This is essentially determined by the ratio
of the concentration of electrons $n$ and the magnetic field $B$.
Therefore, the variation of the concentration of electrons (i.e.,
the chemical potential) is equivalent with variation of the magnetic field in
the quantum Hall system.

In general, the creation of new bands (``gap opening'') can be
described by Dirac fermions \cite{osh,sem,hal}. Starting from the 
tight-binding Hamiltonian the Dirac fermions can be derived in a large
scale approximation. The simplest case is that with half a flux quantum per
lattice plaquette ($\phi=\phi_0/2$) \cite{ludwigetal,fi}. 
(Such a strong magnetic flux is unrealistic in real crystals but typical for
arrays of quantum dots in moderate magnetic fields \cite{weiss}.)
For half a flux quantum per plaquette it is easy to derive the Dirac theory
from a sublattice representation which takes into account the phase factor
$e^{i\pi y/a}$ of the tight-binding Hamiltonian and the staggered potential.
The Fourier components of the non--random part $H(k)$ read
\begin{equation}
{\footnotesize
\pmatrix{
\mu&\!\!\!\!{1+e^{-ik_x}}&\zeta(1-e^{-ik_y})(1-e^{-ik_x})&1+e^{-ik_y}\cr
1+e^{ik_x}&-\mu&1+e^{-ik_y}&-\zeta(1-e^{-ik_y})(1-e^{ik_x})\cr
-\zeta(1-e^{ik_y})(1-e^{ik_x})&1+e^{ik_y}&\mu&-1-e^{ik_x}\cr
1+e^{ik_y}&\zeta(1-e^{ik_y})(1-e^{-ik_x})&-1-e^{-ik_x}&-\mu\cr
}
}
\end{equation}
with $\zeta=it'/4$. All elements of the matrix are measured in units of the
nearest neighbor
hopping rate $t$. After expansion of $k=(\pm\pi,\pm\pi)+ap$ for small $p$
vectors around the four nodes and a global orthogonal transformation 
$H(k)\to OH(k)O$ with
\begin{equation}
O=\pmatrix{
\sigma_0&-i\sigma_0\cr
i\sigma_0&-\sigma_0\cr
}
\end{equation}
the Hamiltonian becomes
\begin{equation}
H'(p)=2\pmatrix{
\mu+4i\zeta&ip_x-p_y&2i\zeta(p_x+p_y)&0\cr
-ip_x-p_y&-\mu-4i\zeta&0&2i\zeta(p_x-p_y)\cr
2i\zeta(p_x+p_y)&0&\mu-4i\zeta&p_y+ip_x\cr
0&2i\zeta(p_x-p_y)&p_y-ip_x&-\mu+4i\zeta\cr
}\equiv\pmatrix{
H_{11}&H_{12}\cr
H_{21}&H_{22}\cr
}.
\end{equation}
The corresponding Green's function
\begin{eqnarray}
& {\hat G}  = \pmatrix{
H_{11}+i\omega&H_{12}\cr
H_{21}&H_{22}+i\omega\cr
}^{-1} &
\nonumber\\
& \sim\pmatrix{
(H_{11}+i\omega)^{-1}&(H_{11}+i\omega)^{-1}H_{12}(H_{22}+i\omega)^{-1}\cr
(H_{22}+i\omega)^{-1}H_{21}(H_{11}+i\omega)^{-1}&(H_{22}+i\omega)^{-1}\cr
}&
\end{eqnarray}
decays asymptotically into two diagonal blocks
\begin{equation}
\sim\pmatrix{
(H_{11}+i\omega)^{-1}&0\cr
0&(H_{22}+i\omega)^{-1}\cr
}.
\end{equation}
The lattice constant $a$ is implicitly scaled out in $t'$, $\mu'$ and $p_j$. 
$\sim$ means asymptotics with respect to $\mu+4i\zeta\sim0$
and $p_x\sim p_y\sim0$.
Thus the approximation breaks up the Hamiltonian (\ref{hamiltonian}) into
two independent Dirac Hamiltonians
$H_{11/22}=\sigma\cdot p+\sigma_3(\mu\mp t')$ with Pauli matrices $\sigma_j$.
The two Dirac theories describe particles with different masses $\mu\mp t'$,
respectively. 
The next nearest neighbor hopping term lifts the degeneracy of the two
Dirac particles. Therefore, it plays an important role in this model and must
be taken into account. Variation of the chemical potential implies a variation
of the Dirac mass. In particular, Dirac fermions undergo a Hall transition if
the mass vanishes \cite{ludwigetal,zie2}. This is a consequence of the fact
that the mass breaks the time-reversal symmetry: depending on the sign of $m$
there is a clockwise or counterclockwise Hall current. If the light Dirac
particle undergoes a Hall transition at $\mu-t'=0$ its contribution to the
Hall conductivity changes from $\sigma_{xy}=-1/2$ for $\mu-t'<0$ to
$\sigma_{xy}=1/2$ for $\mu-t'>0$. (The conductivity is in units of $e^2/h$.)
The heavy Dirac particle contributes $\sigma_{xy}=1/2$ because its mass is
positive. Thus the combined effect is a Hall step from $\sigma_{xy}=0$ to
$\sigma_{xy}=1$. This picture is particularly simple for $\Phi=\Phi_0/2$ but
should also hold for other values of the flux as long as the low energy
excitations are linear and can be described by Dirac fermions.

For large scale properties like for the critical behavior near
the Hall transition it is sufficient to consider only the light particle with
$\mu-t'$
\begin{equation}
(H'+i\omega)^{-1}
=\pmatrix{
i\omega+\mu-t'&i\nabla_1+\nabla_2\cr
i\nabla_1-\nabla_2&i\omega-\mu+t'\cr
}^{-1}
\equiv G(i\omega)\equiv\pmatrix{
G_{11}&G_{12}\cr
G_{21}&G_{22}\cr
},
\label{green}
\end{equation}
where $\nabla$ is the lattice gradient operator.
Disorder, originally introduced in $H$ by the random potential $V$, appears in
$H_{11/22}$ as a diagonal matrix $V'$ with independent random elements
$V_1$, $V_2$. The appearence of two random variables per site is a consequence
of the sublattice representation required by the phase factor $e^{i\pi y/a}$
of the tight-binding Hamiltonian.
The random matrix $V'$ is equivalent to a random mass $\delta\mu\sigma_3$ and a
random energy $\delta E \sigma_0$. For technical reasons the random energy
term will be neglected in the following. 

It should be noticed that the random mass is marginally irrelevant on a
perturbative level \cite{dotsenko}. However, going beyond
perturbation theory, it turns out that the random mass leads to spontaneous
symmetry breaking which changes the properties significantly \cite{zie1}.
This effect has not been included in previous studies of
the localization properties near the Hall transition. It will be important for
the considerations in this article.

Dirac fermions can also be derived as the large scale approximation of the
network model \cite{lee2}. 
It was recently pointed out by Ho and Chalker \cite{ho/chalker} that the
network implies a random Dirac mass (due to fluctuations in the tunneling
rates), a random energy (due to fluctuations in the flux per plaquette) and
a random vector potential (due to fluctuations in the phase of the hopping
elements).
I.e., in terms of the network model the random Dirac mass requires a fixed
flux per plaquette and a fixed phase for the current between the vertices of
the network. This is probably the simplest situation for the realization of a
QHT.

After averaging with respect to the random mass
the localization length $\xi_l$, measured in lattice units 
$a=\sqrt{\phi_0/2B}$ for electrons in a magnetic field $B$,
is defined as the decay length of the function
$C_{jj'}(r,\omega)\equiv\langle|G_{jj'}(r,0;i\omega)|^2\rangle$.
The relation
$|G_{jj'}(r,0;i\omega)|^2=G_{jj'}(r,0;i\omega)G_{j'j}(0,r;-i\omega)$ means
that $\xi_l$ is given by the product of two Green's function at frequencies
with {\it opposite sign} (retarded and advanced Green's functions).
Due to the $2\times 2$ block structure of $G$ there exists a relation between
Green's functions at $i\omega$ and Green's functions at $-i\omega$:
\begin{equation}
G_{jj}(r,r';-i\omega)=-G_{j'j'}(r',r;i\omega),\ \ \
G_{jj'}(r,r';-i\omega)=-G_{jj'}(r',r;i\omega)\ \ \ (j\ne j').
\end{equation}
This identity reflects the Lorentz-covariance of the Dirac theory. It implies 
\begin{equation}
|G_{jj}(r,r';i\omega)|^2$ $=-G_{jj}(r,r';i\omega)G_{j'j'}(r,r';i\omega)
\label{id1}
\end{equation}
and
\begin{equation}
|G_{jj'}(r,r';i\omega)|^2=-G_{jj'}(r,r';i\omega)G_{j'j}(r,r';i\omega).
\label{id2}
\end{equation}
This means that only the Green's functions with one frequency is required for
the evaluation of localization properties in the relativistic
model. The averaged quantity $C_{jj'}(r,\omega)$ is translational invariant.
Therefore, it can be expressed by its Fourier components ${\tilde C}_{jj'}
(k,\omega)$. This can be used to calculate the localization length $\xi_l$.
(The following discussion holds for any choice of $j$, $j'$. Therefore,
these labels are not written explicitly.) The correlation function
$C(r,\omega)$ for large $r$ is proportional to $r^{-\alpha}\exp(-r/\xi_l)$
with some exponent $\alpha$ for which we assume that it is fixed for the
model and does not depend on the parameters. This implies that
\begin{equation}
{\sum_rr^2C(r,\omega)\over\sum_rC(r,\omega)}=
{\sum_rr^{2-\alpha}\exp(-r/\xi_l)\over\sum_rr^{-\alpha}\exp(-r/\xi_l)}
=\xi_l^2{\sum_xx^{2-\alpha}\exp(-x)\over\sum_xx^{-\alpha}\exp(-x)}.
\label{ll}
\end{equation}
Dropping the constant term from the ratio of sums on the right hand side of
(\ref{ll}), the localization length can be defined in terms of the Fourier
components as
\begin{equation}
\xi_l=\sqrt{-{\nabla_k^2{\tilde C}(k,\omega)\over{\tilde C}(k,\omega)}}
\Big|_{k=0}.
\label{local}
\end{equation}
The localization length is finite for $\omega\ne0$ but diverges in the regime
of delocalized states with $\omega\to0$.

The localization length for massless Dirac fermions without disorder
diverges like $|\omega|^{-1}$ if $\omega=0$ is approached. 
This behavior is probably unstable against arbitrarily weak randomness, as it
will be shown in this article. However, it has been shown in a previous paper  
\cite{zie5} that the localization length of the averaged correlation function
has a lower bound which is the energy--energy correlation length of the 2D
random bond Ising model. Since the latter diverges at the critical points,
this implies a divergent localization length at the QHT. 

The calculation for the network model of Chalker and Coddington
\cite{chalk},
indicates that the critical exponent $\nu=1$ of the pure model may change to
$\nu=7/3$ due to
disorder. This was discussed as a possible appearence of
a new random fixed point of the random model \cite{ludwigetal}. However, a 
new fixed point with such a behavior has not been found so far in terms
of renormalization group calculations.

\section{Functional Integral Representation}

It is convenient to introduce a functional integral representation for
$C(r,\omega)$, because this provides a basis to apply an approximation using
a saddle point integration. The product of Green's functions on the right
hand side of (\ref{id1}) and (\ref{id2}) 
can formally be written as
\begin{equation}
(i\omega+H)^{-1}_{rj,r'j'}(i\omega+H^T)^{-1}_{r'k',rk}
=\int\chi_{r'j'}{\bar \chi}_{rj}\Psi_{rk}{\bar \Psi}_{r'k'}\exp(-S_0)
{\cal D}\Psi {\cal D}{\bar \Psi} {\cal D}\chi {\cal D}{\bar \chi}
\label{7}
\end{equation}
with the quadratic form of the superfield $(\chi_r,\Psi_r)$
\begin{equation}
S_0=-i{\rm sign}(\omega)\sum_{r,r'}\pmatrix{
\chi_r\cr
\Psi_r\cr
}\cdot\pmatrix{
i\omega+H&0\cr
0&i\omega+H^T\cr
}_{r,r'}\pmatrix{
{\bar \chi}_{r'}\cr
{\bar \Psi}_{r'}\cr
}.
\end{equation}
$\chi$ is a complex field and $\Psi$ a Grassmann field, respectively. It is
important to notice that $H$ appears in the quadratic form for the
complex field whereas $H^T$ appears for the Grassmann field.
This difference will turn out to be crucial for the localization properties of
the Dirac fermions. In particular, it will give all the delocalized
states expected near the Hall transition. In contrast, the quadratic
form where $H$ is used in the Grassmann sector instead of $H^T$ does not give
these critical properties \cite{zie5,zie2}.

Averaging with respect to disorder leads to
\begin{equation}
\langle(i\omega+H)^{-1}_{rj,r'j'}(i\omega+H^T)^{-1}_{r'k',rk}\rangle
=\int\chi_{r'j'}{\bar \chi}_{rj}\Psi_{rk}{\bar \Psi}_{r'k'}\langle\exp(-S_0)
\rangle {\cal D}\Psi{\cal D}{\bar \Psi}
{\cal D}\chi{\cal D}{\bar \chi}.
\label{8}
\end{equation}
A Gaussian distribution of the random Dirac mass is assumed in the
following with mean $m$ and variance $g$. Then the average can be performed
exactly giving an additional quartic interaction term in $S_0$ with
coupling constant $g$. For weak disorder, i.e., small $g$, one could apply 
perturbation theory. Unfortunately, this does not lead to interesting
results because it can not catch spontaneous symmetry breaking. In order to
deal with the latter one must construct a representation which describes
the field which is the conjugate to the symmetry breaking terms of the
Dirac theory, the mass $m$ and the frequency $\omega$. The appearence
of the symmetry breaking terms in $S_0$ dictates to choose $\chi_r{\bar\chi}_r
$ and $\Psi_r{\bar\Psi}_r$ as the collective fields.

\subsection{Collective Field Representation}

In general, products the fields $\chi,\Psi$ in $S_0$ can be replaced by the 
collective fields as $\chi_r{\bar \chi}_r\to Q_r$,
$\Psi_r{\bar\Psi}_r\to P_r$, $\chi_r{\bar\Psi}_r\to{\bar\Theta}_r$ 
and $\Psi_r{\bar\chi}_r\to\Theta_r$.
(Some care is necessary to choose the paths of integration correctly
\cite{zie3}.) One obtains for (\ref{8}) in collective field representation
(for details see \cite{zie3})
\begin{equation}
{1\over g^2}\int\Theta_{r,kj}{\bar\Theta}_{r',j'k'}\exp(-S'){\cal D} P 
{\cal D}Q{\cal D}\Theta {\cal D}{\bar\Theta}\qquad(r\ne r')
\end{equation}
with the supersymmetric effective action
\begin{eqnarray}
S' & = & {1\over g}\sum_r({\rm Tr}_2Q_r^2+{\rm Tr}_2P_r^2+2{\rm Tr}_2
{\bar \Theta}_r\Theta_r)
\nonumber\\
& & +\ln\det[(H_0+i\omega\sigma_0-2\tau Q\tau)(H_0^T+i\omega\sigma_0+2i\tau
P\tau)^{-1}]
\nonumber\\
& & +\ln\det\big\lbrack{\bf 1}-4\tau{\bar \Theta}\tau(H_0^T+i\omega\sigma_0+2i
\tau P\tau)^{-1}
\tau\Theta\tau (H_0+i\omega\sigma_0-2\tau Q\tau)^{-1}\big\rbrack.
\end{eqnarray}
$H_0=i\sigma\cdot\nabla+m\sigma_3$ is the average Dirac Hamiltonian
and $\tau$ the diagonal matrix $(1,i)$.
The introduction of the collective fields is important to discover the
finite length scale $e^{\pi/g}$, created by disorder, which is crucial
for the properties of the random Dirac mass model \cite{zie3}.


\subsection{Saddle Point Approximation}

A saddle point (SP) approximation is a crude approach for a two--dimensional
system because it usually gives wrong results for low dimensional systems
due to strong effects of
fluctuations. In the model under consideration it will be used as a
starting point to study also the fluctuations. An argument in favor of
a SP approach is the fact that some features of the model can
be described which are not available from pertubation theory. An example
is the creation of states in the massive Dirac theory due to randomness
\cite{zie1,zie1a}. The hope is that the fluctuations around the SP
are controlled by Gaussian fluctuations, at least if randomness is weak.
This will be supported by the discussion presented below. Another argument
for the SP approximation is its equivalence with the
$N\to\infty$--limit, where $N$ is a formal extension of the model to one
for electrons with $N$ states per lattice site \cite{zie2}.

The SP of the functional integral is given by the equations
$\delta_QS'=\delta_PS'=0$.
The two SP equations are identical if one substitutes $P=iQ$:
\begin{equation}
\sigma_3\tau Q_r\tau\sigma_3 =g (H_0+i\omega\sigma_0-2\tau Q\tau)^{-1}_{r,r}.
\end{equation}
An ansatz for a uniform SP solution reads $\tau Q_0\tau =-(i\eta\sigma_0+m_s
\sigma_3)/2$.
The SP equations imply a shift of the frequency $\omega\to\eta'\equiv\eta+
\omega$ with
\begin{equation}
\eta'-\omega =\eta' g I
\label{spea}
\end{equation}
and a shift of the average mass $m\to m'\equiv m+m_s$ with
\begin{equation}
m_s=-mgI/(1+gI),
\end{equation}
\begin{equation}
I\sim {1 \over \pi}\int_0^1({\eta'}^2+(m+m_s)^2+k^2)^{-1}kdk
\sim -{1\over 2\pi}\ln\lbrack {\eta'}^2 +(m+m_s)^2\rbrack=
-{1\over \pi}\ln|\mu|
\end{equation}
with $\mu=m+m_s+i\eta'$. 
In the pure limit $g\to0$ the SP equations lead to $\eta'=\omega$ and
$m_s=0$. For a given $g>0$ the SP depends on two parameters, $m$ and $\omega$.
For $\omega=0$ and large $|m|$ there is only a trivial solution of (\ref{spea}) 
with $\eta'=\eta=0$ because $gI<1$.
As one varies $|m|$ there is a critical point $m_c=2\mu_c=2e^{-\pi/g}$
where $gI<1$ approaches $gI=1$. As a consequence,
the SP solution of (\ref{spea}) bifurcates from $\eta=0$ to $\eta\ne0$
at $\mu=\mu_c$, and $\eta=0$ becomes unstable \cite{zie3}. In the following
only the region with $\eta\ne0$ is considered, where $sign(\eta)
=sign(\omega)$. This has a non--zero density of states
\begin{equation}
\rho(m)\approx(1/2\pi g)\lim_{\omega\to0}\eta'=(1/2\pi g)\eta=
(1/4\pi g)\sqrt{(m_c^2-m^2)}\Theta(m_c^2-m^2).
\end{equation}
which describes a semicircular behavior. Of course, this must be normalized
with an energy cut--off dependent constant.

As an ansatz for the SP approximation with $\omega\ne 0$ one can write
\begin{equation}
\eta'(\omega)^2=\eta'(\omega=0)^2+\delta={m_c^2-m^2\over4}+\delta,
\label{3.11}
\end{equation}
where the last equation follows from the SP equation for $\omega=0$. This
implies
\begin{equation}
|\mu|^2=m_c^2/4+\delta
\end{equation}
and for eq. (\ref{spea})
\begin{equation}
(g/2\pi)\eta'\ln(1+4\delta/m_c^2)=\omega.
\label{spe}
\end{equation}
The expansion of the logarithm for $\delta\ll m_c^2/4$ yields a cubic
equation for $\delta$
\begin{equation}
({m_c^2-m^2\over4}+\delta)\delta^2\approx\omega^2({\pi m_c^2\over2g})^2.
\end{equation}
Although this equation could be solved directly it is simpler to
distinguish two different asymptotic regimes:

\noindent
(i) $(m_c^2-m^2)/4\ll\delta\ll m_c^2/4$:
\begin{equation}
\delta\sim\omega^{2/3}({\pi m_c^2\over2g})^{2/3}
\label{spe1}
\end{equation}
and

\noindent
(ii) $(m_c^2-m^2)/4\gg\delta$:
\begin{equation}
\delta\sim\omega{\pi m_c^2\over g}
{1\over\sqrt{m_c^2-m^2}}.
\label{spe2}
\end{equation}


\subsection{Gaussian Fluctuations}

In order to evaluate the localization length $\xi_l$ the Gaussian fluctuations
around the SP must be calculated (semi-classical approximation).
Since $Q$, $P$ and $\Theta$ are $2\times2$ matrices, the fluctuations can also
be written as 4-component vector fields:
$q_1=\delta Q_{11}$, $q_2=(\delta Q_{12}+\delta
Q_{21})/2$, $q_3=-i(\delta Q_{12}-\delta Q_{21})/2$, $q_4=\delta Q_{22}$ with
analogous definitions for $p_1,...,p_4$ and for the Grassmann field
$\psi_1,...,\psi_4$.
The action of the Gaussian fluctuations reads in Fourier representation
\begin{equation}S'\approx \int\sum_{\mu,\mu'=1}^4 [({\bf I}_k)_{\mu,\mu'}
(q_{k,\mu}q_{-k,\mu'}+p_{k,\mu}p_{-k,\mu'})+2({\bf I}_k')_{\mu,\mu'}
{\bar \psi}_{k,\mu}\psi_{-k,\mu'}]d^2k
\end{equation}
with the fluctuation matrices ${\bf I}_k$ and ${\bf I}'_k$. For the large
scale properties one needs only the asymptotic behavior for small
wave vectors $k$. In particular, for a vanishing wave vector there is
\begin{equation}
{\bf I}_0=\pmatrix{
1/g-2\alpha\mu^{*2}&0&0&2\beta\cr
0&2/g-4\alpha|\mu|^2&0&0\cr
0&0&2/g-4\alpha|\mu|^2&0\cr
2\beta&0&0&1/g-2\alpha\mu^2\cr
}
\end{equation}
\begin{equation}
{\bf I'}_0=\pmatrix{
1/g-2\alpha\mu^{*2}&0&0&0\cr
0&2(1/g-2\alpha|\mu|^2-2\beta)&0&0\cr
0&0&2(1/g-2\alpha|\mu|^2+2\beta)&0\cr
0&0&0&1/g-2\alpha\mu^2\cr
}
\end{equation}
with
\begin{equation}
\alpha=\int(|\mu|^2+k^2)^{-2}d^2k/4\pi^2\sim |\mu|^{-2}/4\pi=1/\pi m_c^2
\end{equation}
and
\begin{equation}
\beta=\int k^2(|\mu|^2+k^2)^{-2}d^2k/4\pi^2=I/2-1/4\pi
\sim -\ln|\mu|^2/4\pi\sim1/2g.
\end{equation}
These quantities become quite large for small $|\mu|$ indicating a short
range behavior of the related modes. However, the second diagonal element
of ${\bf I'}$ vanishes for vanishing $\omega$ because of
$1/g-2\alpha|\mu|^2-2\beta=\omega/g\eta'$. This is a direct consequence of the
SP condition (\ref{spea}) and indicates a critical mode $\psi_{k,2}$ for
all $|m|< m_c$. Moreover, it implies a divergent behavior of the
localization length if $\omega\to0$. The corresponding correlation
function of the critical mode can be calculated in the large scale limit
by expanding ${\bf I'}_k$ in powers of $k$ as
\begin{equation}
{\tilde C}(k,\omega)=g^{-2}(2\omega/g\eta'+Dk^2)^{-1},
\label{cricor}
\end{equation}
where
\begin{equation}
D=4\alpha\Big[
1+\alpha({\mu^2\over1/g-2\alpha\mu^2}+{\mu^{*2}\over1/g-2\alpha
\mu^{*2}})\Big].
\label{3.21}
\end{equation}
$g\eta'D/2$ is like a diffusion coefficient. It is real and it never becomes
zero.
The critical behavior describes a phase with a divergent sum
$\sum_{r,j}\langle |G_{jj}(r,r')|^2\rangle$; i.e., the correlation 
function decays non--exponentially. It implies that 
${\tilde C}(k=0,\omega)=\eta'/2g\omega=\pi\langle\rho(\omega)\rangle/
\omega$. This holds not only on the SP level but in general due to the
identity $\sum_{r',j,j'}|G_{jj'}(r,r')|^2=\pi\rho /\omega$.

${\tilde C}$ in (\ref{cricor}) is an approximation based on $Dk^2\ll1$.
Since $D$ diverges like $1/\pi m_c^2$ it is not possible to use this
expression for the pure limit. This also reflects the non-perturbative
character of the SP approximation.

Apart from the critical (delocalized) fermion mode there is a boson mode
which becomes critical at $m=\pm m_c$. This is due to a vanishing eigenvalue 
of ${\bf I}_0$ because $\eta=0$ at $m=\pm m_c$. Thus there are delocalized
states for $|m|<m_c$ due to massless fermions whereas a combination of
critical fermions {\it and} critical bosons controls the QHT. The band of 
delocalized fermions simplifies the study of transport properties of the 
model away from the critical points $m=\pm m_c$.

An analogous calculation for the Gaussian fluctuations can be performed for
Anderson localization (i.e., for $\phi=0$ in eq. (\ref{hamiltonian})). In
that case the Green's function of eq. (\ref{green}) must be replaced by
\begin{equation}
G(i\omega)\equiv\pmatrix{
i\omega+M+\nabla^2&0\cr
0&i\omega-M-\nabla^2\cr
}^{-1}.
\end{equation}
Now the matrix $H=M+\nabla^2$ is symmetric in contrast to 
$H=i\sigma\cdot\nabla+(\mu-t')\sigma_3$. As a consequence the corresponding
fluctuation matrix is degenerated for the fermion and the boson sector
(i.e., ${\bf I}'_k={\bf I}_k$) with
\begin{equation}
{\bf I}_0=\pmatrix{
1/g-\alpha&0&0&0\cr
0&2(1/g-\beta)&0&0\cr
0&0&2(1/g-\beta)&0\cr
0&0&0&1/g-\alpha^*\cr
},
\end{equation}
where
\begin{equation}\alpha={1\over (2\pi )^d}\int {(k^2-E+i\eta)^2\over
((k^2-E)^2+\eta^2)^2}d^dk
\end{equation}
and  
\begin{equation}\beta={1\over (2\pi )^d}\int {1\over (k^2-E)^2+\eta^2}d^dk.
\end{equation}
There are two vanishing eigenvalues of ${\bf I}_0$ for vanishing $\omega$ in
the Grassmann as well as in the complex sector, each of them is
$1/g-\beta=\omega/g\eta'$. Therefore, the large scale behavior of this model
is very different, and the critical properties are described by a nonlinear
sigma model including fermion as well as boson degrees of freedom
\cite{efetov}. This is a consequence of the fact that 
$\phi=0$ preserves the supersymmetry whereas the model with $\phi=\phi_0/2$
breaks the supersymmetry implying ${\bf I}\ne{\bf I'}$. The latter implies
that there is only one massless (delocalized) fermion field, and all
other fields are massive (localized).

\subsection{Corrections to Gaussian Fluctuations}

Gaussian fluctuations are usually not sufficient to describe the properties
of a critical system, especially at low dimensionality, because interaction
between the fluctuations are a relevant perturbation. For Anderson 
localization the interaction is marginal in $d=2$ as one finds from power
counting of the scaling behavior. This method can also be applied to the
effective field theory of random Dirac fermions. It provides a first
check for the effect of the interaction among the fluctuations.
In the following we will see that the perturbation term for random Dirac
fermions has dimensionality $-2$. Therefore, the interaction of the
fluctuations is irrelevant. (It seems that the dimensionality is reduced by
2 in comparison to Anderson localization. This is similar to the
dimensional reduction by 2 in supersymmetric theories, applied to the
average density of states for a particle in a strong magnetic field and
a random potential \cite{wegner,brezinetal}.)

Away from the critical points $|m|=m_c$ it is sufficient to study the 
Grassmann fluctuations because the complex fluctuations are massive. Their
action reads for terms up to fourth order in the fluctuations
\begin{equation}S'\approx 2\int
(2\omega/g\eta'+Dk^2){\bar\psi}_{k,2}\psi_{-k,2}d^2k
-8\sum_{r_1,...,r_4}
Tr_2[{\bar\Theta}_{r_1}{\bar G}_{r_1r_2}\Theta_{r_2}{\bar G}^T_{r_2r_3}
{\bar\Theta}_{r_3}{\bar G}_{r_3r_4}\Theta_{r_4}{\bar G}^T_{r_4r_1}],
\end{equation}
where ${\bar G}=\tau G_0\tau$.
$\psi_2\sigma_1$ is the only critical mode of the collective Grassmann
field $\Theta_r$. Thus the interaction term can also be written as
\begin{equation}
-8\sum_{r_1,...,r_4}
Tr_2[\sigma_1{\bar G}_{r_1r_2}\sigma_1{\bar G}^T_{r_2r_3}\sigma_1
{\bar G}_{r_3r_4}\sigma_1{\bar G}^T_{r_4r_1}]
{\bar\psi}_{r_1,2}\psi_{r_2,2}{\bar\psi}_{r_3,2}\psi_{r_4,2}.
\label{int}
\end{equation}
The trace term can be evaluated and yields after some straightforward
calculation together with the approximation $G_{r,r'}\approx 
|\mu|^{-2}[(i\eta' \sigma_0-m\sigma_3)\delta_{r,r'}+i\nabla_{1;r,r'}
\sigma_1+i\nabla_{2;r,r'}\sigma_2]$ the following expression
\begin{equation}
-8|\mu|^{-8}\sum_{r_1,...,r_4}\big[
{\bar\psi}_{r_1}d_{r_1,r_2}\psi_{r_2}d_{r_2,r_3}{\bar\psi}_{r_3}d_{r_3,r_4}
\psi_{r_4}d_{r_4,r_1}+
{\bar\psi}_{r_1}d_{r_1,r_2}^*\psi_{r_2}d_{r_2,r_3}^*{\bar\psi}_{r_3}
d_{r_3,r_4}^*\psi_{r_4}d_{r_4,r_1}]
\end{equation}
where the index 2 of the Grassmann field has been dropped and $d_{r,r'}=
(\nabla_{1;r,r'}+i\nabla_{2;r,r'})$. This result
reflects the fact that terms with an odd number of $\nabla$--operators
cancels in (\ref{int}) and terms quadratic in $\nabla$ cancel each other
because of the anticommutation rule of the Grassmann field:
${\bar\psi}_{r_1}\psi_{r_1}{\bar\psi}_{r_3}\psi_{r_3}+
{\bar\psi}_{r_1}\psi_{r_3}{\bar\psi}_{r_3}\psi_{r_1}$=0. Higher order
terms with at most second order gradients disappear individually because
they contain Grassmann fields at the same site.
Simple power counting indicates that this interaction term has dimensionality
$-2$. Therefore, the interaction is irrelevant in comparison with Gaussian
fluctuations considered in the previous section, and it scales quickly to 
zero under renormalization group transformations.

The situation is different if we approach $m=\pm m_c$ because the complex
field also becomes critical. As a consequence, the corrections
to Gaussian fluctuations are marginal rather than irrelevant then.
It is possible that the localization length becomes finite at $m=\pm m_c$ due
to renormalization effects. This phenomenon near $\pm m_c$ requires a separate
treatment which will not be considered in this article.

The irrelevance of the interaction terms is very special for the model under
consideration. In similar two-dimensional models, like the Gross-Neveu
model or the tight binding model without magnetic field (orthogonal nonlinear
sigma  model) or with weak magnetic field (unitary nonlinear sigma model),
the interaction is always marginal (i.e., dimensionality of the interaction
term is zero).

\section{Localization Length}

According to the discussion of the Gaussian fluctuations in Sect. IIIC
the leading large scale behavior is given by the correlation function
of (\ref{cricor}). From the latter the localization length can be
calculated, using eq. (\ref{local}), as $\xi_l=(2Dg\eta'/\omega)^{1/2}$.
Together with the SP results (\ref{3.11}) and (\ref{3.21}) $\xi_l$ reads
\begin{equation}
\xi_l\approx\sqrt{2g\over\pi}\Big({m_c^2-m^2\over4}+\delta\Big)^{1/4}
(m_c^2/4+\delta)^{-1/2}\omega^{-1/2}.
\label{fig1}
\end{equation}
The localization length diverges like $\omega^{-1/2}$ because $D$ and $\eta'$
remain non-zero for $|m|<m_c$ according to the results of the SP approximation.
To compare $\xi_l$ with numerical or experimental results it is important
that $D$ can be large ($\sim |\mu|^{-2}$) and $\eta'$ can be small.
Therefore, several quasi--scaling regimes exist as indicated by the
graph in Fig. 1 which is simply the plot of (\ref{fig1}) together with
(\ref{spe}). Only in the asymptotic regime $\omega\sim0$, i.e., for 
$(m_c^2-m^2)/4\gg\delta$, the localization length diverges like 
$\omega^{-1/2}$ 
\begin{equation}
\xi_l\sim\sqrt{2g\over\pi}\Big({m_c^2-m^2\over4}\Big)^{1/4}{2\over m_c}
\omega^{-1/2}.
\end{equation}
Surprisingly, the exponent $\nu=1/2$ agrees with that of a completely
different approach to the QHT by Affleck \cite{affleck}. Moreover, $\nu=1/2$
was also found by Ludwig et al. for Dirac fermions with a random vector
potential at that special point on their fixed point line where the average
density of states is finite \cite{ludwigetal}.

A quasi--scaling regime occurs for $(m_c^2-m^2)/4\ll\delta\ll m_c^2/4$
where we have
\begin{equation}
\xi_l\sim\sqrt{2g\over\pi}(m_c^2/4+\delta)^{-1/2}
({\pi m_c^2\over2g})^{1/6}\omega^{-1/3}\sim
\sqrt{2g\over\pi}{2\over m_c}({\pi m_c^2\over2g})^{1/6}\omega^{-1/3},
\label{coeff1}
\end{equation}
i.e., the effective exponent is $\nu=1/3$.

Going back to the SP equation (\ref{spe}) one could also assume 
$m_c^2/4\ll\delta$. It gives
\begin{equation}
-\delta^{1/2}(g/2\pi)\ln(m_c^2)=\delta^{1/2}\sim\omega.
\label{spe3}
\end{equation}
This result implies for the localization length
\begin{equation}
\xi_l\sim\sqrt{2g\over\pi}\delta^{-1/4}\omega^{-1/2}\sim \sqrt{2g\over\pi}
\omega^{-1},
\label{coeff2}
\end{equation}
i.e., the effective exponent is $\nu=1$.
The coefficient in (\ref{coeff1}) {\it decays} for small $g$ like
$g^{1/3}e^{2\pi/3g}$ whereas
the corresponding coefficient in (\ref{coeff2}) {\it grows} like $g^{1/2}$
with the strength of disorder. The coefficients of the power law for
$\nu=1/3$ ($c_1$), for $\nu=1/2$ ($c_2$) and for $\nu=1$ ($c_3$) are
plotted in Fig. 2.
$\nu=1$ and the behavior of the corresponding coefficient $c_3$ agree
with the observation in experiments \cite{dol1,dol2,sha}.

\section{Conductivity}

The longitudinal conductivity can be calculated via Kubo's formula
\begin{equation}
\sigma_{xx}(\omega)={e^2\over h}\omega^2
\sum_rr^2\langle G_{jj'}(r,0;i\omega)G_{j'j}(0,r;-i\omega)\rangle.
\end{equation}
The correlation function is again the expression we have considered in the
effective field theory. Since for small $\omega$ only the large scale part of
the correlation contributes significantly, we can use
${\tilde C}(k,\omega)$ of (\ref{cricor}) to write
\begin{equation}
\sigma_{xx}(\omega)=
-{e^2\over h}\omega^2\nabla_k^2{\tilde C}(k,\omega)|_{k=0}=
{e^2\over h}\omega^2g^{-2} D(\omega/g\eta')^{-2}={e^2\over h}D\eta'^2.
\end{equation}
For weak disorder we use the approximation $D\sim4\alpha$.
Furthermore, for $\omega\sim0$ we get $\alpha=1/4\pi |\mu|^2\sim1/\pi m_c^2$.
Therefore, the conductivity reads in the dc--limit
\begin{equation}
\sigma_{xx}(\omega=0)\sim{e^2\over\pi h}{m_c^2-m^2\over m_c^2}
\Theta(m_c^2-m^2).
\end{equation}
For $m=0$ the dc--conductivity is independent of disorder
\begin{equation}
\sigma_{xx}^c(\omega=0)\sim e^2/\pi h.
\end{equation}
This result is in agreement with a calculation for Dirac fermions with a
random vector potential\cite{ludwigetal}. Thus there is a universal
conductivity in $d=2$ at the center of the band of extended states
$-m_c<m<m_c$. 

The Hall conductivity can be calculated using a simpler field theory
which does not break the supersymmetry \cite{zie2}. In units of $e^2/h$
it was found
\begin{equation}
\sigma_{xy}\approx1/2+{\rm sign}(m)\Big[1/2-(1/\pi)
{\rm arctan}(\sqrt{m_c^2/m^2-1})\Theta(m_c^2-m^2)\Big].
\end{equation}
For $m=0$ the Hall conductivity is always $e^2/2h$ whereas for $m\ne0$ it
depends
on disorder via $m_c$. The resistivity, calculated from the averaged
conductivities (which is an additional approximation because one should
actually evaluate the averaged resistivity) yields 
\begin{equation}
\rho_{xx}^c=\sigma_{xx}^c/({\sigma_{xx}^c}^2+{\sigma_{xy}^c}^2)
\approx 0.9 h/e^2.
\end{equation}
This value agrees with an error of about $\pm10\%$ with various
experimental results \cite{shaharetal}.

\section{Discussion of the Results}

For a given strength of randomness $g$ there are three different regimes of
the behavior of the localization length, depending on the value of $\delta$.
$\delta$ is directly related to the frequency $\omega$ according to the
equations (\ref{spe1}), (\ref{spe2}) and (\ref{spe3}), respectively. The
corresponding power laws $\xi_l\sim c_j\omega^{-\nu_j}$ appear with effective
exponents $\nu_1=1/3$ and $\nu_3=1$ and with the asymptotic exponent
$\nu_2=1/2$.
The exponents and the corresponding coefficients $c_j$ are shown in Table I,
and the $g$ dependence of the coefficients is plotted in Fig. 2.  
Ludwig et al. evaluated the exponent $\nu$ for two-dimensional Dirac fermions
with random vector potential with variance $g_a$. In that case the exponent
of the asymptotic localization length depends smoothly on randomness as
$\nu=1/(1+g_a/\pi)$ \cite{ludwigetal}. It should be noticed that the inequality $\nu\ge1$ of
Chayes et al. \cite{chayes} for two--dimensional random systems does not
apply to the problem under consideration (cf also the discussion in
Ref.\cite{ludwigetal}). It is remarkable though that there is agreement of the
asymptotic result $\nu=1/2$ between the present calculation for the random
Dirac mass, the random vector potential with finite average density of states
(where $g_a/\pi=1$) 
\cite{ludwigetal} and the nonlinear sigma model with topological term
\cite{affleck}.

The result $\nu=1/3$ is not reliable because it appears close to the critical
points $|m|=m_c$, where the bosonic degrees of freedom become critical.
Moreover, the asymptotic regime with $\nu=1/2$ is not realistic:  
The typical width of the fluctuations $\sqrt{g}$ in a sample is about
$10\%$ of the hopping rate. I.e., $m_c=2e^{-\pi/g}$
is immeasurably small. Therefore, only the regime $\delta\gg m_c^2/4$ would be
accessible, and only the effective exponent $\nu=1$, together
with the coefficient $\sqrt{2g/\pi}$, is of practical relevance.
Consequently, only the coefficient of the localization length is affected
by randomness. This is in agreement with the observations in 
Refs.\cite{dol1,dol2,sha}. 
However, the divergency of the localization length is controversal among
the various experiments. In this context it would be interesting if the
cross-over length, evaluated in the article as $\sim\exp(\pi/g)$, can be
observed experimentally.

\noindent
The density of states is non--zero near the QHT. This result is 
non--perturbative because $\rho\propto \exp(-\pi/g)$, and it reflects
spontaneous symmetry breaking \cite{zie1}. It agrees with a
Monte Carlo simulation for the network model by Lee and Wang \cite{lee/wang},
with an exact diagonalization of a finite system \cite{hatsugai}, with
a rigorous estimation \cite{zie1} and with an exact calculation for 
Lorentzian disorder \cite{zie1a}.

\noindent
The longitudinal conductivity $\sigma_{xx}(\omega)$ is non--zero between
$m=-m_c$ and $m=m_c$, the transition region between two Hall plateaux.
In an experiment it may not be possible to resolve the width of this band of
delocalized states
since $m_c$ is too small. Therefore, the width of the transition between the
Hall plateaux will always be dominated by thermal broadening. Thus a power
law for the width $\Delta$ at temperature $T$ like $\Delta\sim T^\kappa$
\cite{wei} is a realistic ansatz.
The conductivity at the QHT should agree with the value
of $\sigma_{xx}$ at $m=0$, i.e., $\sigma_{xx}^c=e^2/h\pi$. Converting this
value into the corresponding value of $\rho_{xx}^c$ gives $\approx 0.9h/e^2$
which is in good agreement
with experiments. The universal value $\sigma_{xx}^c=e^2/h\pi$ agrees with
the value found for random vector potential \cite{ludwigetal}
but not with the result found
from the numerical simulation of the network model, where $\sigma_{xx}^c=
e^2/2h$ was found \cite{lee}. This value was also obtained for the lowest
Landau level projection \cite{gammel}.

It seems that our results for the 2D Dirac fermions with random mass are in
good agreement with the corresponding results for 2D Dirac fermions with
random vector potential and with recent experiments. But there is disagreement
with the results of the numerical simulation of the network model. This may
be related to different types of randomness (e.g., strong randomness in the
magnetic field) or to different geometries.
It is also possible that there is a strong renormalization of the localization
length near $m=\pm m_c$ due to extended boson fields leading to an
exponent $\nu\approx 7/3$. This, however, would raise the question about the
origin of the experimental value $\nu\approx1$. Moreover, the disagreement
of the values of $\sigma_{xx}^c$ for the network model on the one side and for
Dirac fermions with $m=0$ on the other side cannot be explained by 
renormalization effects.

\begin{table}
\caption{Exponents for the localization length and the corresponding
coefficients.}
\begin{tabular}{c c c c}
regime & $\delta\ll(m_c^2-m^2)/4$ & $(m_c^2-m^2)/4\ll\delta\ll m_c^2/4$ 
& $\delta\gg m_c^2/4$\\
\tableline
$\nu$ & $1/2$ & $1/3$ &  $1$ \\
coefficient & $c_2$ & $c_1$ &  $c_3$
\end{tabular}
\end{table}

\begin{figure}
\caption{Scaling of the localization length as a function of the frequency
for disorder strength $g=0.2$ and $\mu-t'=0$.}
\caption{Coefficients $c_j$ of the power laws of $\xi_l$ for $\mu-t'=0$.}
\end{figure}

\begin{references}

\bibitem{koch}
S. Koch, R.J. Haug, K. von Klitzing and K. Ploog, Phys.Rev. B {\bf 46}, 1596
(1992)

\bibitem{dol1}
V.T. Dolgopolov, G.V. Kravchenko and A.A. Shashkin, Phys.Rev. B {\bf 46},
13303 (1992)

\bibitem{dol2}
A.A. Shashkin, V.T. Dolgopolov and G.V. Kravchenko, Phys.Rev. B {\bf 49}, 14486
(1994)

\bibitem{sha}
A.A. Shashkin, V.T. Dolgopolov, G.V. Kravchenko, M. Wendel, R. Schuster,
J.P. Kotthaus, R.J. Haug, K. von Klitzing, K. Ploog, H. Nickel and W. Schlapp,
Phys.Rev.Lett. {\bf 73}, 3141 (1994)

\bibitem{LR}
P.A. Lee and T.V. Ramakrishnan, Rev.Mod.Phys. {\bf 57}, 287 (1985)

\bibitem{chalk}
J.T. Chalker and P.D. Coddington, J.Phys. C {\bf 21}, 2665 (1988)

\bibitem{lee}
D.H. Lee, Z.Wang and S.Kivelson, Phys.Rev.Lett {\bf 70}, 4130 (1993)

\bibitem{gammel}
B.M. Gammel and W. Brenig, Phys.Rev.Lett {\bf 73}, 3286 (1984)

\bibitem{huck}
B. Huckestein and B. Kramer, Phys.Rev.Lett. {\bf 64}, 1437 (1990)

\bibitem{huo}
Y. Huo and R.N. Bhatt, Phys.Rev.Lett. {\bf 68}, 1375 (1992)

\bibitem{affleck}
I. Affleck, Nucl.Phys. B {\bf 265} [FS15], 409 (1986)

\bibitem{pruisken}
H. Levine, S.B. Libby and A.M.M. Pruisken, Phys.Rev.Lett. {\bf 51}, 1915
(1983);
A.M.M.Pruisken, in {\sl The Quantum Hall Effect}, edited by R.E.Prange and
S.M.Girvin (Springer-Verlag, New York, 1990)

\bibitem{ludwigetal}
A.W.W. Ludwig, M.P.A. Fisher, R. Shankar and G. Grinstein, Phys. Rev. 
B {\bf 50}, 7526 (1994)

\bibitem{thou}
D.J. Thouless, M. Kohmoto, M.P. Nightingale and M. den Nijs, Phys.Rev.Lett.
{\bf 49}, 405 (1982)

\bibitem{hats}
Y. Hatsugai, Phys.Rev. B {\bf 48}, 11851 (1993)

\bibitem{osh}
M. Oshikawa, Phys.Rev. B {\bf 50}, 17357 (1994)

\bibitem{sem}
G.W. Semenoff, Phys.Rev.Lett. {\bf 53}, 2449 (1984)

\bibitem{hal}
F.D.M. Haldane, Phys.Rev.Lett. {\bf 61}, 2015 (1988)

\bibitem{fi}
M.P.A. Fisher and E. Fradkin, Nucl.Phys. B {\bf 251} [FS13], 457 (1985)

\bibitem{weiss}
D. Weiss, {\sl Quantum Dynamics of Submicron Structures}, edited by H.A.
Cerdeira et al. (Kluwer Academic Publishers, 1995)

\bibitem{zie2}
K. Ziegler, Europhys.Lett. {\bf 28}, 49 (1994)

\bibitem{dotsenko}
Vi. Dotsenko and Vl.S. Dotsenko, Adv.Phys.{\bf 32}, 129 (1983)

\bibitem{zie1}
K. Ziegler, Nucl.Phys. B {\bf 285} [FS19], 606 (1987)

\bibitem{lee2}
D.H. Lee, Phys.Rev. B {\bf 50}, 10788 (1994)

\bibitem{ho/chalker}
C.-M. Ho and J.T. Chalker, preprint cond-mat/9605073

\bibitem{zie5}
K. Ziegler, Europhys.Lett. {\bf 31}, 549 (1995)


\bibitem{zie3}
K. Ziegler, Nucl.Phys. B {\bf 344}, 499 (1990)

\bibitem{zie1a}
K. Ziegler, Phys.Rev. B {\bf 53}, 9653 (1996)


\bibitem{efetov}
K.B.Efetov, Adv.Phys. {\bf 32}, 53 (1983)

\bibitem{wegner}
F. Wegner, Z.Phys. B {\bf 51}, 279 (1983)

\bibitem{brezinetal}
E. Br\'ezin, D. Gross and C. Itzykson, Nucl.Phys. B {\bf 235} [FS11], 24 (1984)


\bibitem{shaharetal}
D. Shahar, D.C. Tsui, M. Shayegan and R.N. Bhatt, Phys.Rev.Lett. {\bf 74},
4511 (1995)

\bibitem{chayes}
J.T. Chayes, L. Chayes, D.S. Fisher and T. Spencer, Phys.Rev.Lett. {\bf 57},
2999 (1986)

\bibitem{lee/wang}
D.-H. Lee and Z. Wang,  Phil.Mag.Lett. {\bf 73}, 145 (1996) 

\bibitem{hatsugai}
Y. Hatsugai and P.A. Lee, Phys.Rev. {\bf B48}, 4204 (1993)

\bibitem{wei}
H.P. Wei, D.C. Tsui, M.A. Paalanen and A.M.M. Pruisken, Phys.Rev.Lett.
{\bf 61}, 1294 (1988) 

\end{references}
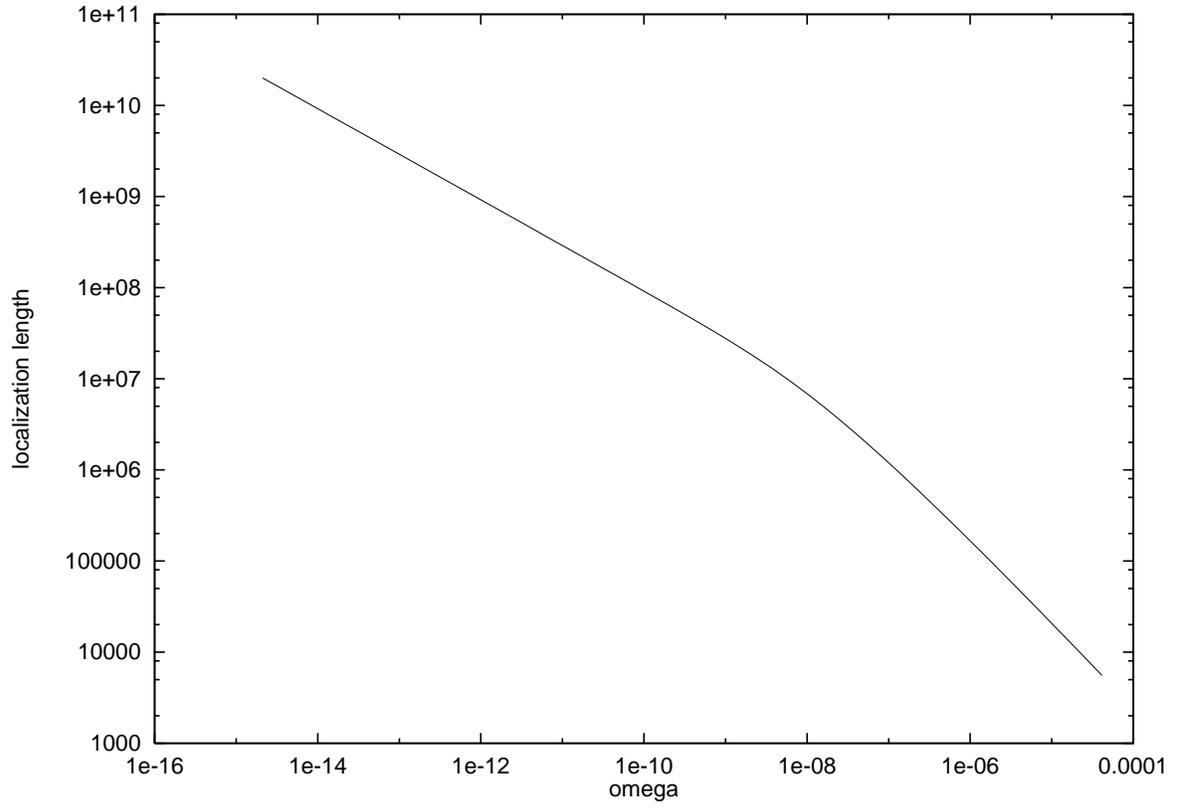
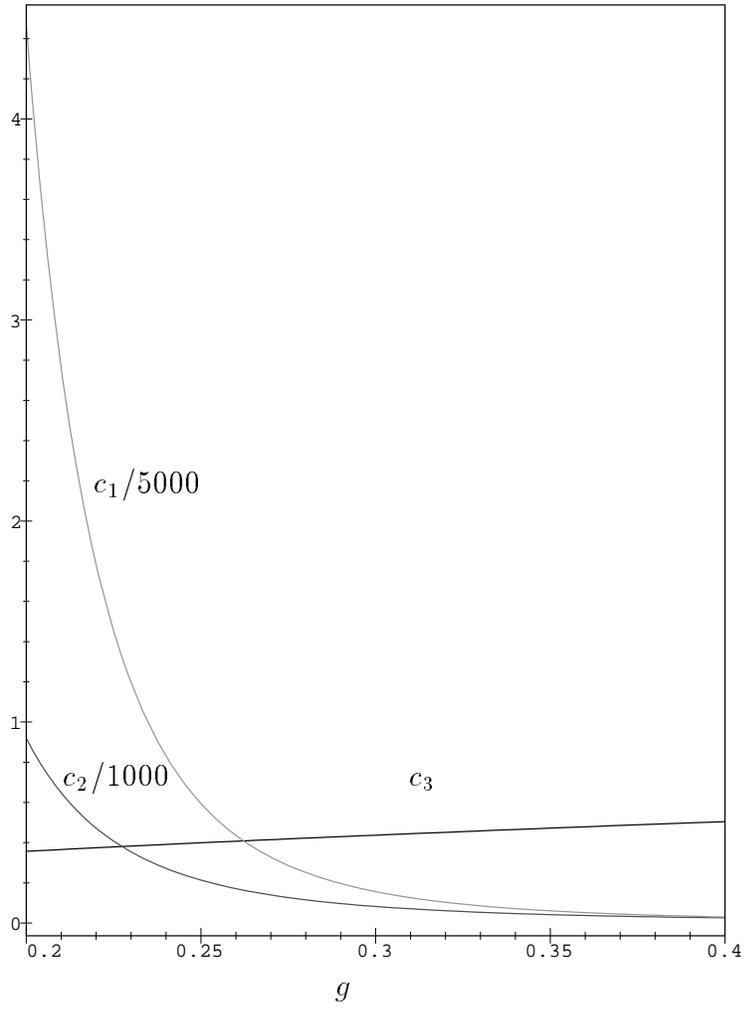
\end{document}